\documentclass[epj]{svjour}
\usepackage{graphics}
\usepackage{epstopdf}

\usepackage{amsmath}
\usepackage{graphics}
\usepackage{flushend}
\usepackage{cuted}
\usepackage{bm}% bold math
\usepackage{widetext}
\usepackage{subfigure}
\usepackage[dvips]{graphicx}
\usepackage{color}
\usepackage{sidecap}
\newcommand{\bsig}{{\bm\sigma}}

\newcommand{\hbrho}{\hat{{\bm\rho}}}
\newcommand{\hrho}{\hat{\rho}}

\newcommand{\be}{\begin{equation}}
\newcommand{\ee}{\end{equation}}

\newcommand{\bea}{\begin{eqnarray}}
\newcommand{\eea}{\end{eqnarray}}
\newcommand{\bd}{\begin{displaymath}}
\newcommand{\ed}{\end{displaymath}}
\newcommand{\ba}{\begin{array}}
\newcommand{\ea}{\end{array}}
\newcommand{\bit}{\begin{itemize}}
\newcommand{\ei}{\end{itemize}}
\newcommand{\bc}{\begin{center}}
\newcommand{\ec}{\end{center}}
\newcommand{\bfl}{\begin{flushleft}}
\newcommand{\efl}{\end{flushleft}}
\newcommand{\bfr}{\begin{flushright}}
\newcommand{\efr}{\end{flushright}}

\newcommand{\s}{\c{s}}

\newcommand{\hatt}{\hat{t}}
\newcommand{\hG}{\hat{G}}
\newcommand{\hV}{\hat{V}}
\newcommand{\tilt}{\tilde{t}}
\newcommand{\tL}{\tilde{\Lambda}}

\newcommand{\om}{i\omega_n}

\newcommand{\CP}{CePt$_3$Si}

\def\ua{\uparrow}
\def\da{\downarrow}

\def\br{{\bf r}}
\def\bk{{\bf k}} \def\bq{{\bf q}} 
 
\def\hg{\rm\hat{g}}
\def\i{\rm i}
\def\bg{{\bf g}} 
  
\def\hbg{\hat{{\bf g}}} \def\bd{{\bf d}}  \def\bS{{\bf S}}
 \def\bS{{\bf S}} \def\hbz{\hat{{\bf z}}}
\def\hbx{\hat{{\bf x}}} \def\hby{\hat{{\bf y}}} 

\def\da{\downarrow} \def\ua{\uparrow} 

\def\6{\partial}

\def\e{\epsilon}

\def\s{\sigma} \def\t{\tau}

\def\={\!\!\!&=&\!\!\!}
\def\+{\!\!\!&&\!\!\!+~}
\def\-{\!\!\!&&\!\!\!-~}

\begin{document}
\title{Full t-matrix approach to quasiparticle interference in non-centrosymmetric superconductors}
%\subtitle{Full t-matrix approach to QPI in non-centrosymmetric superconductors}
\author{Alireza Akbari\inst{1,2} and  Peter Thalmeier\inst{1}% etc
% \thanks is optional - remove next line if not needed
%\thanks{\emph{Present address:} Insert the address here if needed}%
}                     % Do not remove
%\offprints{}          % Insert a name or remove this line
%
\institute{
Max Planck Institute for Chemical Physics of Solids, D-01187 Dresden, Germany 
\and 
Max Planck Institute for Solid State Research, D-70569 Stuttgart, Germany
}
\date{\today } %/ Revised version: date}
% The correct dates will be entered by Springer
%
\abstract{
 We develop the full t-matrix theory of quasiparticle interference (QPI) for non-centrosymmetric (NCS) superconductors with
 Rashba spin-orbit coupling. We give a closed solution for the QPI spectrum for arbitrary combination and strength of
 nonmagnetic ($V_c$) and magnetic ($V_m$) impurity scattering potentials in terms of integrated normal and anomalous Green's functions. The theory is applied to a realistic 2D model of the Ce-based 131- type heavy fermion superconductors.
We discuss the QPI dependence on frequency, composition and strength of scattering and compare with Born approximation results.
We show that the QPI pattern is remarkably stable against changes in the scattering model and can therefore give reliable information on the properties of Rashba-split Fermi surface sheets and in particular on the accidental nodal position of the mixed singlet-triplet gap function in NCS superconductors.
 % end of PACS codes
} %end of abstract
\titlerunning{Full t-matrix approach to QPI in non-centrosymmetric superconductors}
\authorrunning{ Akbari and   Thalmeier }% etc

\maketitle

%========================================================
%========================================================
%========================================================
%========================================================
%========================================================
%========================================================
%========================================================
%========================================================

%%%%%%%%%%%%%%%%%%%%%%%%

\section{Introduction}
\label{sec:introduction}

The determination of gap symmetry and nodal positions is the most important problem in unconventional superconductors (SC).
Various methods are available that give at least a partial knowledge on the gap properties. Particular useful methods for this purpose are angle-resoved photoemission spectroscopy (ARPES) experiments \cite{damascelli:03,okazaki:12}, specific heat and thermal transport measurements
in a rotating field \cite{matsuda:06} which are based on the Doppler-shift effect.  More recently STM-based quasiparticle interference technique (QPI) has been applied which utilizes the ripples in electronic density generated by random magnetic or nonmagnetic  surface impurities. 
 In the normal state they are determined only by the shape of the Fermi surface sheets if the impurity scattering is isotropic. However, in the superconducting state the opening of an anisotropic gap introduces typical changes in the shape of the QPI spectrum from which information on the gap symmetry may be obtained. Experimentally this method has been used, e.g. in cuprates \cite{McElroy:2003,Hanaguri:2009} and Fe-pnictides \cite{Hanaguri:2010,allan:12}. Theoretical investigations were given in \cite{byers:93,capriotti:03,pereg:03,Wang:2003,zhu:04,nunner:06,Maltseva:2009,andersen:09} for the cuprates and 
\cite{Akbari:2010,Knolle:2010,huang:11} for the pnictide systems.  In these high-T$_c$ superconductors, however, ARPES  is also readily applicable for determination of gap anisotropy.  Sofar this is not possible for heavy fermion unconventional superconductors where SC gaps are only in the range of ~ 1 meV.  In this case the former two methods are more powerful. QPI technique has recently been proposed as a way to discriminate between different d-wave pairing states in 115 systems \cite{akbari:11} and was for the first time successfully demonstrated for CeCoIn$_5$ \cite{allan:13,zhou:13}. Before it was also used to investigate  quasiparticle properties in the hidden order state of URu$_2$Si$_2$ \cite{aynajian:10,schmidt:10,yuan:12}.%\\

In the NCS superconductors with inversion symmetry breaking the nodal positions are accidental and not determined by symmetry.
Examples are the tetragonal heavy fermion 131 and 113 compounds \cite{settai:07} like CePt$_3$Si \cite{bauer:07} and CeRhSi$_3$ \cite{kimura:07}. QPI might provide a new way to determine their position and the singlet-triplet mixture of the order parameter in these compounds more precisely. 
No experimental results have been reported but the theory in weak scattering Born approximation was developed for NCS superconductors with Rashba spin orbit coupling \cite{akbari:13}. 
It was shown that a number  of unconventional QPI features are to be expected: Pronounced differences in the charge- and spin- QPI appear due to the effect of Rashba coherence factors. The latter exhibits Rashba - induced anisotropies even for isotropic exchange scattering. Furthermore a new kind of cross- QPI between charge and spin channel appears which is directly related to the non-zero Rashba vector.%\\

It remained however unclear to which extend the conclusions on QPI pattern from Born approximation are stable with respect to the frequency dependence of the t-matrix that inevitably appears for stronger scattering, giving even the possibility of resonance formation.
Therefore in this work we develop the full t-matrix QPI theory for NCS superconductors without posing any conditions on the absolute and relative strengths of nonmagnetic ($V_c$) and magnetic ($V_m$) impurity scattering. 
Unlike in most treatments we give the results of full t-matrix theory as closed analytical expressions where only the momentum integration of Green's functions has to be performed numerically. We show that in full t-matrix theory beyond Born approximation two new aspects appear: i) The QPI functions are the sum of diagonal (non-spin flip) and anti-diagonal (spin flip) part. The latter are present only in the superconducting phase and correspond to Andreev scattering terms that change  particles into holes and vice versa.  
ii) Due to the imaginary part of the full t-matrix a new integration kernel appears in the diagonal part which is not present in Born approximation. We will show that nevertheless the momentum-space pattern of  QPI function is remarkably stable under the change of potential model or whether Born approximation of full t-matrix theory is used. Therefore our results strengthen confidence that QPI investigation could give reliable results on the accidental node positions in NCS Rashba superconductors, just as it was able to identify the symmetry-determined node positions of the $d_{x^2-y^2}$ singlet superconductor
\break
CeCoIn$_5$ \cite{allan:13}.

%%%%%%%%%%%%%%%%%%%%%% figure %%%%%%%%%%%%%%%%%%%%%%%%%%%%%%%%%%%%%%%%%%%%%
\begin{figure}%[t]
\centerline{
\hspace{-1.61cm}
\includegraphics[width=0.56\textwidth]{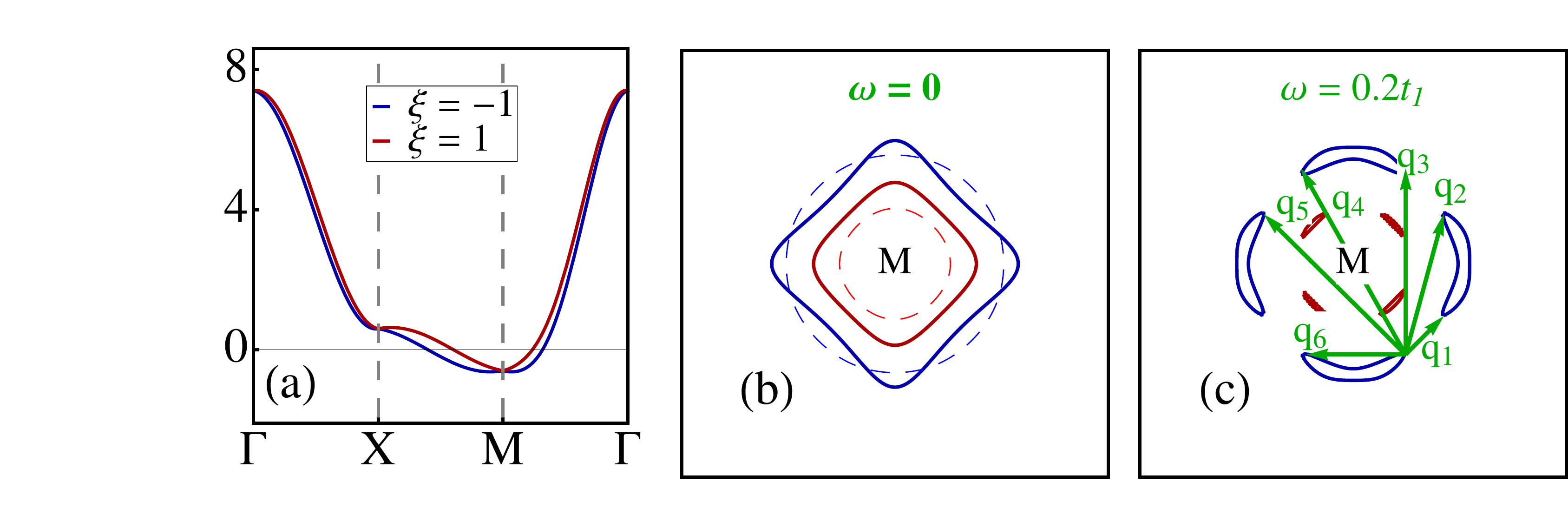}
}
\caption{ (Color online) 
a)
2D band structure along ${\rm \Gamma XM \Gamma}$ with $ t_2=0.35t_1, g=0.2t_1$, and $\mu=-2t_1$.
The total band width is  $W\simeq 8t_1 \equiv T^*\simeq 14$ K (exp. Kondo temperature \cite{bauer:07}) 
$\equiv 1.2$ meV corresponding to reference energy scale $t_1=0.15$ meV.
b) Normal state electron Fermi surface around M$(\pi,\pi)$ point. Dashed lines indicate nodes of gap functions $\Delta_{\bk\xi}$ with parameters: $\psi_0=2t_1$, $\psi_1=t_1$, and $\phi_0=0.65t_1$ (also in subsequent figures). Only $\Delta_{\bk -}$ (blue) has nodes on the Fermi surface.
c)  Spectral function for the superconducting state around M point.
Set of characteristic QPI wave vectors $(\bq_1-\bq_6)$ correspond to those in  the spectral 
functions given in Figs. \ref{Fig4}-\ref{Fig6}.
 Momentum range in b) and c) is given by $-\pi\leq (q_{x,y}-\pi) \leq\pi$.
}\vspace{-0.31cm}
%}
\label{Fig1}
\end{figure}
%%%%%%%%%%%%%%%%%%%%%%%%%%%%% figure %%%%%%%%%%%%%%%%%%%%%%%%%%%%%%%%%%%%%%%

%
%%%%%%%%%%%%%%%%%%%%%% figure %%%%%%%%%%%%%%%%%%%%%%%%%%%%%%%%%%%%%%%%%%%%%
\begin{figure*}%[t]
\centerline{
%\hspace{-1.61cm}
\includegraphics[width=1.18\linewidth]{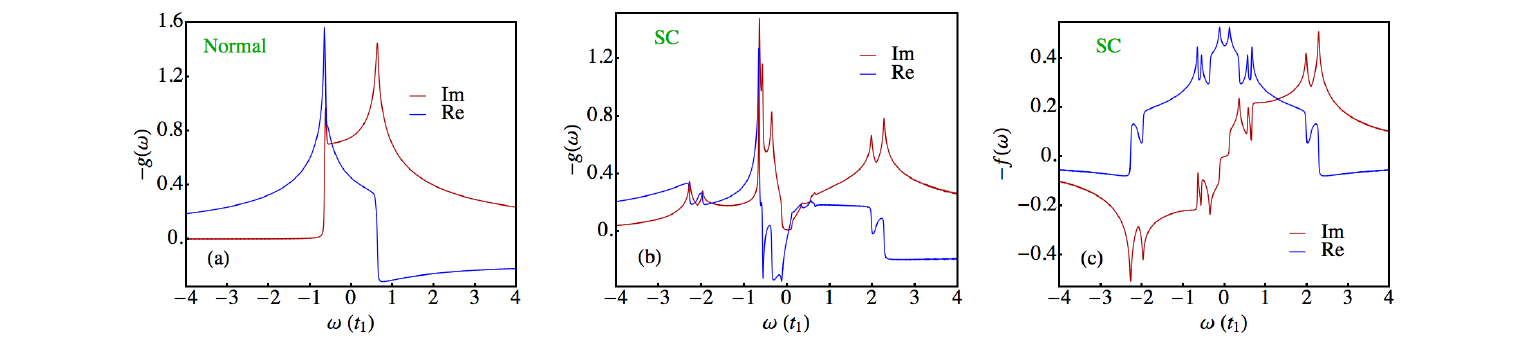}
}
\caption{ (Color online) 
a) Real and imaginary parts of   $g(\omega)$ in the normal state versus $\omega$.
b)  Real and imaginary parts of   $g(\omega)$ 
in the superconducting state versus $\omega$.
c) Real and imaginary parts of   $f(\omega)$  in the superconducting state versus $\omega$. 
In the normal state $f(\omega)\equiv 0$.
}\vspace{-0.31cm}
%}
\label{Fig2}
\end{figure*}
%%%%%%%%%%%%%%%%%%%%%%%%%%%%% figure %%%%%%%%%%%%%%%%%%%%%%%%%%%%%%%%%%%%%%%

\section{Model of electronic states in non-centrosymmetric compounds}
\label{sec:model}

The BCS model for non-centrosymmetric superconductors is given by \cite{frigeri:06,eremin:06,fujimoto:07}
\bea
\hspace{-1cm}
H_{SC}&=&\sum_{\bk\s\s'}
\bigg[(\varepsilon_\bk-\mu)\s_0+\bg_\bk\cdot\bsig\bigg]_{\s\s'}c_{\bk\s}^\dagger c_{\bk\s'}
\nonumber\\
&&+
\frac{1}{2}\sum_{\bk\s\s'}(\Delta_\bk^{\s\s'}c_{-\bk\s}^\dagger c_{\bk\s'}^\dagger +H.c.),
\label{eq:HBCS}
\eea
where $\varepsilon_\bk$ is the band energy $\mu$ the chemical potential.
Furthermore $\bg_\bk=-\bg_{-\bk}$ defines the antisymmetric Rashba spin orbit coupling term due to broken inversion symmetry \cite{samokhin:04}. Therefore the superconducting $2\times 2$ gap matrix  $\Delta_\bk=[\psi_\bk\s_0+\bd_\bk\cdot\bsig]i\s_y$ has even singlet ($\psi_\bk$) as well as odd triplet ($\bd_\bk$) components. The latter
must be parallel to the Rashba vector $\bd_\bk=\phi_\bk\bg_\bk$ to avoid detrimental effects by pairbreaking \cite{sigrist:09}. Here the fully symmetric $\phi_\bk\equiv \phi_0$ is set to a constant. Furthermore $\bsig =(\sigma_x,\sigma_y,\sigma_z)$ denotes the Pauli matrices. After diagonalization of the model we obtain an effective two band superconductor on the Rashba split $(\xi=\pm1)$ bands given by $\e_{\bk\xi}=\varepsilon_\bk-\mu+\xi |\bg_\bk|$.  They correspond to split Fermi surface (FS) sheets ($\e_{\bk\xi}=0$) with opposite helical spin polarizations and two different superconducting gaps $\Delta_{\bk\xi}=\psi_\bk+\xi |\bd_\bk|$.%\\

We will investigate in detail a 2D model for Ce- based 131 systems \cite{takimoto:08,takimoto:09}. The possible effects of magnetic order in these compounds \cite{yanase:08} are not treated here. For the kinetic energy we consider only in-plane dispersion given by
 $\varepsilon_\bk=2t_1(\cos k_x+\cos k_y)+4t_2 \cos k_x \cos k_y$, where $t_1$ and $t_2$ are nearest and next nearest neighbor hopping, respectively.
Furthermore we choose 
\be
\bg_\bk=g(\sin k_y\hbx-\sin k_x\hby)=g(\sin k_y,-\sin k_x, 0),
\label{eq:rashba}
\ee
in the tetragonal plane \cite{sigrist:09}. The resulting Rashba-split bands 
 are shown in Fig.~\ref{Fig1}a and the electron-type Fermi surface around the M$(\pi,\pi)$ points in Fig.~\ref{Fig1}b. The split constant-energy surfaces ($\omega > 0$) for the superconducting  state are presented in Fig.~\ref{Fig1}c. It is known from thermal conductivity \cite{izawa:05} that the superconducting gap has line nodes (in 3D). To achieve nodes on the M-point Fermi surface we use an extended s-wave \cite{sigrist:09}  form for $\psi_\bk$ with $A_{1g}$ (full) symmetry and as before $\bd_\bk=\phi_0\bg_\bk$. The latter belongs to the nontrivial $A_{2u}$ triplet representation of $C_{4v}$ symmetry group that transforms like the $k_y\hbx-k_x\hby$ basis function: It is invariant under $2C_2$ and $C_4$ rotations but changes sign under $2\sigma_v$ and $2\sigma_d$ reflections from mirror planes parallel to the tetragonal axes and diagonals, respectively \cite{sigrist:96,mineev:99}. For the quasiparticle states of $H_{SC}$ this leads to the two distinct gaps $(\xi=\pm 1)$
\be
\Delta_{\bk\xi}=\psi_0+\psi_1(\cos k_x+\cos k_y)+\xi \phi_0g \sqrt{\sin^2k_x+\sin^2k_y},
\label{eq:xigap}
\ee
on the Rashba-split Fermi surfaces $\epsilon_{\bk\xi}=0$. The corresponding quasiparticle energies are given by  $E_{\bk\xi}=[\epsilon_{\bk\xi}^2+\Delta_{\bk\xi}^2]^\frac{1}{2}$. 
The gap zeroes of the hybrid gap function are accidental and their existence requires the fine-tuning of singlet and triplet amplitudes $\psi_1$ and $\phi_0$ in Eq.~(\ref{eq:xigap}) (see caption of Fig.~\ref{Fig1}). For positive parameters  the nodes (node lines in 3D) appear only for the $E_{\bk -}$ quasiparticle  sheet but not for $E_{\bk +}$  and are shown as dashed lines in Fig.~\ref{Fig1}b. The evolution of constant energy surfaces in the SC state is shown in Fig.~\ref{Fig1}c. A few wave vectors $\bq_i$ connecting high curvature points close to nodal positions that will appear prominently in QPI spectra for small $\omega$ are indicated by arrows. For larger $\omega$  FS sheets are reconnected.
For the calculation of the QPI pattern we use the normal and anomalous $2\times 2$  spin-space matrix Green's functions of the unperturbed system
\be
\begin{aligned}
\label{eq:greenform}
G(\bk,i\omega_n)&=G_+(\bk,i\omega_n)\s_0+G_-(\bk,i\omega_n)(\hbg_\bk\cdot\bsig),
%\ 
\\
F(\bk,i\omega_n)&=[F_+(\bk,i\omega_n)\s_0+F_-(\bk,i\omega_n)(\hbg_\bk\cdot\bsig)]i\sigma_y,
\end{aligned}
\ee
where the unit Rashba vector is defined by $\hbg_\bk=\bg_\bk/|\bg_\bk|$. The scalar bare Green's functions are given by
\begin{equation}
\begin{aligned}
\label{eq:green}
&%&
G_+(\bk,i\omega_n)=\frac{1}{2}\sum_\xi 
\frac{
i\omega_n+\e_{\bk\xi}}
{(i\omega_n)^2-E^2_{\bk\xi}},
\\&
G_-(\bk,i\omega_n)=\frac{1}{2}\sum_\xi 
\xi
\frac{
i\omega_n+\e_{\bk\xi}}
{(i\omega_n)^2-E^2_{\bk\xi}},
\\&F_+(\bk,i\omega_n)=
\frac{1}{2}\sum_\xi 
\frac{
\Delta_{\bk\xi}
}{
(i\omega_n)^2-E^2_{\bk\xi}
},
\\&%&
F_-(\bk,i\omega_n)=
\frac{1}{2}\sum_\xi 
\xi
\frac{
\Delta_{\bk\xi}
}{
(i\omega_n)^2-E^2_{\bk\xi}
}.
\end{aligned}
\end{equation}

\section{Scattering potential}
\label{sect:scattering}

The QPI density oscillations are obtained from the full Green's function
that  is determined by the effect of  scattering from random charge and spin impurities at the surface. We treat both cases and express the total scattering Hamiltonian containing nonmagnetic potential and isotropic exchange scattering in compact form as 
\bea
{\it H}_{imp}=\sum_{\bk\bq\alpha}V_\alpha(\bq)S_\alpha\Psi_{\bk+\bq}^\dagger\hat{\rho}_\alpha\Psi_\bk.
\label{eq:impurity}
\eea
Here we use the Nambu 4-component spinor representation $\Psi^\dagger=(c^\dagger_{\bk\ua}, c^\dagger_{\bk\da}, c_{-\bk\ua}, c_{-\bk\da})$. In the Nambu space the $4\times4$ matrices $\hrho_\alpha$ ($\alpha =(0,{\i}) =(0,x,y,z)$) are given by \cite{maki:69}  $\{\hrho_\alpha\}=(\hrho_0,\hbrho) = (\tau_3\sigma_0,\tau_0\sigma_x,\tau_3\sigma_y,\tau_0\sigma_z)$. Here the $\tau_\alpha$- and $\sigma_\alpha$- Pauli matrices (with $\tau_0=\sigma_0\equiv \mbox{I}$) act on Nambu and spin indices, respectively. Furthermore we define $\{S_\alpha\}=(1,\bS)$. The first index $\alpha=0$ corresponds to nonmagnetic impurity scattering $V_0(\bq)$ and entries ${\i}=x,y,z$ to  isotropic magnetic exchange scattering $V_{\i}(\bq)=V_{ex}(\bq)$  from impurity spins \bS. The spin components $S_{\i}$ are assumed as frozen, i.e. polarized in a given direction by a small magnetic field. An important consequence of the Rashba coupling is that spin and charge channels for impurity scattering are not decoupled as in the case of centrosymmetric metals where $\bg_\bk = 0$. This has also been shown for spin and charge response functions \cite{takimoto:08}.
 The \bq-dependence of the scattering is due to two effects \cite{capriotti:03,zhu:04}, the finite extension of the potential at the impurity site and the (random) distribution of the impurities. The former tends to cut off the QPI spectrum at large momenta and the latter causes its blurring. These are side effects which will not be considered further, therefore in the following we will restrict to momentum independent $V_\alpha$.
 
 %
%%%%%%%%%%%%%%%%%%%%%% figure %%%%%%%%%%%%%%%%%%%%%%%%%%%%%%%%%%%%%%%%%%%%%
\begin{figure*}%[t]
\centerline{
\includegraphics[width=0.88\linewidth]{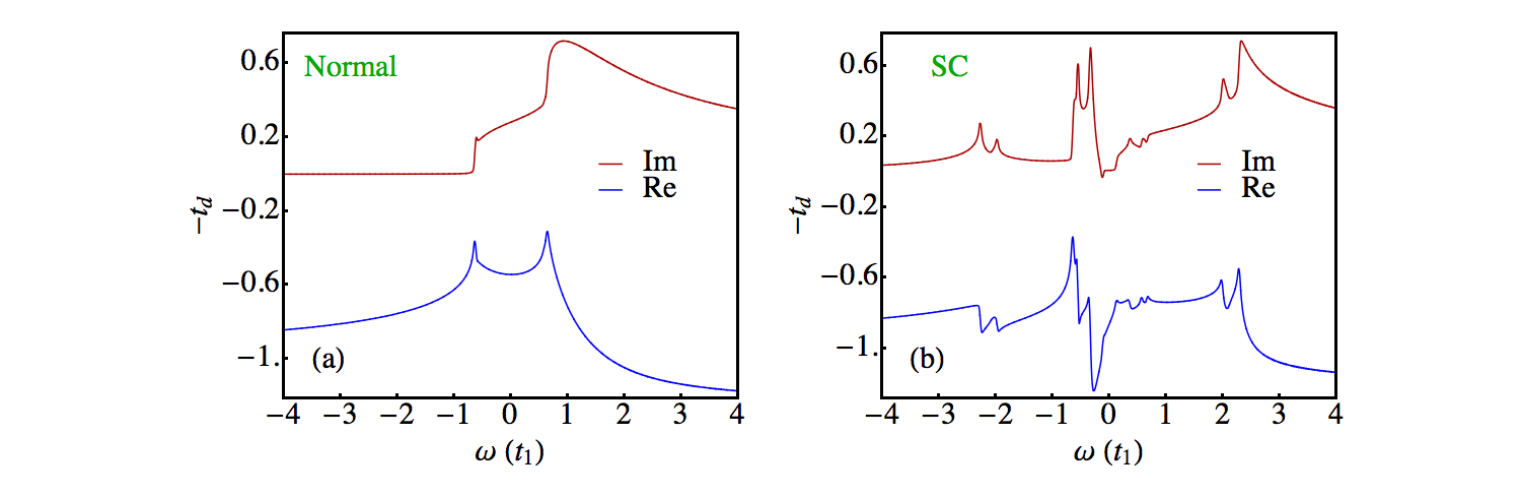}
}
\caption{ (Color online) 
a)   Real and imaginary parts of diagonal t-matrix element  $t_d(\omega)$ 
in the normal state versus $\omega$.
b)   Real and imaginary parts of   $t_d(\omega)$ 
in the superconducting state versus $\omega$. Here $V_c =  t_1$ and $V_m =0$.
}\vspace{-0.31cm}
%}
\label{Fig3}
\end{figure*}
%%%%%%%%%%%%%%%%%%%%%%%%%%%%% figure %%%%%%%%%%%%%%%%%%%%%%%%%%%%%%%%%%%%%%%

\section{T-matrix theory}
\label{sec:tmatrix}

The t-matrix describing the repeated scattering of conduction electrons at an impurity site is given by the Lippmann-Schwinger equation
in Nambu space according to
\be
\hatt_{\bk\bk'}(\om)=\hV_{\bk\bk'}+\sum_{\bk''}\hV_{\bk\bk''}\hG_{\bk''}(\om)\hatt_{\bk''\bk'}(\om).
\label{eq:LS}
\ee
Where $\hV_{\bk\bk'}=\hV(\bq)$ with momentum transfer $\bq=\bk-\bk'$ is the impurity scattering potential 
given in Eq.~(\ref{eq:impurity}).
The above equation can only be solved in closed form under the assumption that $\hV(\bq)$ is independent of the momentum transfer.
This corresponds to on-site impurity scattering potentials only. We restrict to the case that the impurity spin is polarized in z-direction (up or down) by a small external field. Then
\be
\hV=V_c\hrho_0+V_m\hrho_z = V_c\t_3\s_0+V_m\t_0\s_z,
\label{eq:scatt}
\ee
with $V_c=V_0$ and $V_m=S_zV_z$ giving the strength of nonmagnetic (charge) and magnetic (spin exchange) scattering, respectively. Then the momentum independent t-matrix is given by
\be
\hatt(\om)=[1-\hV{\hg}(\om)]^{-1}\hV.
\label{eq:tmat}
\ee
Here ${\hg}(\om)=(1/N)\sum_{\bk}\hG(\bk,\om)$ is the momentum integral over the $4\times 4$ Green's function matrix 
\be
\hat{G}( {\bf  k},\om)=
\left[
 \begin{array}{cc}
 G( {\bf  k},\om) & F( {\bf  k},\om) \\
 F^\dagger( {\bf  k},\om) &  -G(-{\bf  k},-\om) 
\end{array}
\right],
\label{eq:greeenmat}
\ee
whose entries are the $2\times 2$ normal and and anomalous Green's functions in spin space according to Eq.~(\ref{eq:greenform}).
Because $G_\xi(\bk,\om)$ and  $F_\xi(\bk,\om)$ are symmetric and  the Rashba vector $\hbg_\bk$ is antisymmetric under inversion $\bk\rightarrow -\bk$ one always has 
$$\frac{1}{N}\sum_{\bk}G_-(\bk,\om)\hbg_\bk =0,$$
 and  
 $$\frac{1}{N}\sum_{\bk}F_-(\bk,\om)\hbg_\bk =0.$$ 
 This simplifies our analysis considerably because then only two integrated scalar Green's functions 
 $$g(\om)\equiv g_+(\om)=\frac{1}{N}\sum_{\bk}G_+(\bk,\om),$$ 
 and
  $$f(\om)\equiv f_+(\om)=\frac{1}{N}\sum_{\bk}F_+(\bk,\om)$$ remain. Explicitly we have
\be
\begin{aligned}
g(i\omega_n)&=\frac{1}{2N}\sum_{\bk\xi} 
\frac{(i\omega_n+\e_{\bk\xi})}{[(i\omega_n)^2-E^2_{\bk\xi}]},
\\
f(i\omega_n)&=\frac{1}{2N}\sum_{\bk\xi} 
\frac{\Delta_{\bk\xi}}{[(i\omega_n)^2-E^2_{\bk\xi}]}.
\label{eq:greenint}
\end{aligned}
\ee
These function have the symmetry properties
$g(\om)^*=g(-\om)$ and $f(\om)^*=f(\om)=f(-\om)$.
Then the integrated $4\times 4$ - matrix Green's function can be written as
\be
\hat{g}(\om)=
\left[
 \begin{array}{cc}
 g(\om)\sigma_0 & f(\om)(i\sigma_y) \\
 f(\om)^*(i\sigma_y)^\dagger &  -g(\om)^*\sigma_0
\end{array}
\right].
\label{eq:gintmat}
\ee
Using this result and the expression for the scattering potential in Eq.~(\ref{eq:scatt}) we obtain the final result for the $4\times 4$ t- matrix by inversion as 
\begin{widetext}
\be
\hat{t}(\om)=
\left[
 \begin{array}{cc}
t_d^*(\om)\sigma_0+\tilt_d^*(\om)\sigma_z & -\tilt_a(\om)\sigma_x-t_a(\om)(i\sigma_y) \\
 -\tilt_a(\om)\sigma_x-t_a(\om)(i\sigma_y)^\dagger &  -t_d(\om)\sigma_0+\tilt_d(\om)\sigma_z 
\end{array}
\right].
\label{eq:tmat1}
\ee
\end{widetext}
The diagonal (d) and antidigaonal (a) t-matrix elements are explicitly given by
\bea
\begin{aligned}
t_d(i\omega_n)&=\frac{1}{2}\sum_{\nu=\pm}\frac
{V_\nu[1-V_\nu g(i\omega_n)]}
{|1-V_\nu g(i\omega_n)|^2+V_\nu^2 f(i\omega_n)^2},
\\
\tilt_d(i\omega_n)&=\frac{1}{2}\sum_{\nu=\pm}\nu\frac
{V_\nu[1-V_\nu g(i\omega_n)]}
{|1-V_\nu g(i\omega_n)|^2+V_\nu^2 f(i\omega_n)^2},
\\
t_a(i\omega_n)&=\frac{1}{2}\sum_{\nu=\pm}\frac
{V_\nu^2f(i\omega_n)}
{|1-V_\nu g(i\omega_n)|^2+V_\nu^2 f(i\omega_n)^2},
\\
\tilt_a(i\omega_n)&=\frac{1}{2}\sum_{\nu=\pm}\nu\frac
{V_\nu^2f(i\omega_n)}
{|1-V_\nu g(i\omega_n)|^2+V_\nu^2 f(i\omega_n)^2}.
%\\
\label{eq:telement}
\end{aligned}
\eea
Here $t_d(i\omega_n)^*=t_d(-i\omega_n)$, 
$\tilt_d(i\omega_n)^*=\tilt_d(-i\omega_n)$, 
\break
 $t_a(i\omega_n)^*=t_a(i\omega_n)$, 
 and 
 $\tilt_a(i\omega_n)^*=\tilt_a(i\omega_n)$.
 Furthermore we defined $\nu=\pm$ and  $V_\pm=V_c\pm V_m$ as the sum or difference of normal and
magnetic scattering. In the special case of only nonmagnetic scattering $(V_m=0, V_\nu =V_c)$ the matrix elements simplify to
\be
\begin{aligned}
&
t_d(i\omega_n)=\frac
{V_c[1-V_c g(i\omega_n)]}
{|1-V_cg(i\omega_n)|^2+V_c^2 f(i\omega_n)^2}, %\;\;\;\;
\\
&
t_a(i\omega_n)=\frac
{V_c^2f(i\omega_n)}
{|1-V_cg(i\omega_n)|^2+V_c^2 f(i\omega_n)^2},%\;\;\;\;
\label{eq:telement1}
\end{aligned}
\ee
and $\tilt_d(i\omega_n)=\tilt_a(i\omega_n)=0$. In this case the t-matrix reduces to
\bea
\hat{t}(\om)=
\left[
 \begin{array}{cc}
t_d^*(\om)\sigma_0& -t_a(\om)(i\sigma_y) \\
-t_a(\om)(i\sigma_y)^\dagger &  -t_d(\om)\sigma_0
\end{array}
\right],
\label{eq:tmat2}
\eea
which has the same spin-space structure as ${\hg}(\om)$  in Eq.~(\ref{eq:gintmat}) due to the scalar scattering potential $V_c$.\\

The anti-diagonal (a) blocks in the t-matrices   Eqs.~(\ref{eq:tmat1},\ref{eq:tmat2}) which change the Nambu pseudo-spin $\tau_z$ appear only in the superconducting state where $f(\om) \neq 0$. They correspond to an Andreev-type scattering process where the holes scatter to electron states and vice versa due to the presence of the condensate. They are absent in Born approximation where no intermediate anomalous Green's function $F(\bk,\om)$ appears in the scattering. Since the Bogoliubov quasiparticle states in the presence of the condensate are superpositions of up-spin electrons and down-spin holes the Andreev -type scattering will also lead to a spin-flip as seen from the general case in Eq.~(\ref{eq:tmat1}), even for the case of a pure scalar scattering potential in  Eq.~(\ref{eq:tmat2}).
Using the above closed analytical solution for the t-matrix we can now calculate the QPI spectrum.

%%%%%%%%%%%%%%%%%%%%%% figure %%%%%%%%%%%%%%%%%%%%%%%%%%%%%%%%%%%%%%%%%%%%%
\begin{figure}
  \centering
  \includegraphics[width=0.49\textwidth]{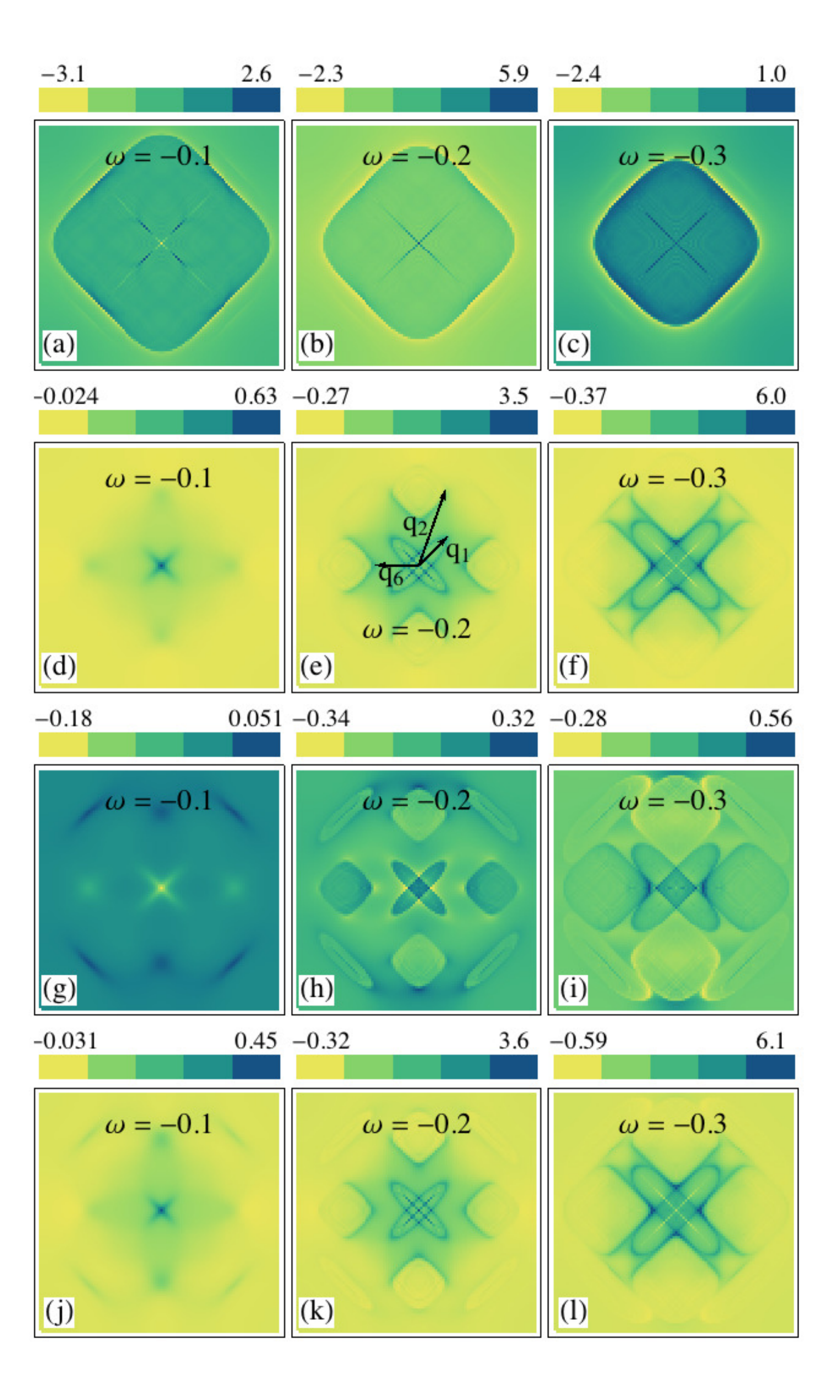}% picture filename
  \caption{ (Color online) 
a-c) Total charge- QPI (${\tilde\Lambda}_{0}(\bq,\omega)$) in the normal state for different $\omega<0$ ($t_1$ units) and  scattering from non-magnetic impurities.
j-l) The same quantity for the superconducting state.
d-f) represent the diagonal contributions  (${\tilde\Lambda}_{0}^d$) and  g-i) represent the anti-diagonal contributions   (${\tilde\Lambda}_{0}^a$)  of the total charge- QPI for the superconducting state.
The frequency $\omega =V$ satisfies $\omega < |\Delta_{\bk\xi}|\simeq t_1 \ll W = T^*=8t_1$.
Here $V_c =t_1$ and $V_m = 0$.  The momentum range in each panel is given by $-\pi\leq q_{x,y}\leq\pi$. This applies also to all following figures.
}
\label{Fig4}
\end{figure}
%%%%%%%%%%%%%%%%%%%%%%fig%%%%%%%%%%%%%%%%%%%%%%%%%%%%%%%%%%%%%%%%%%%%%

\section{QPI conductance}
\label{sec:QPI}

We calculate the change in STM tunneling conductance in charge or spin channel $\alpha (0,x,y,z) $ due to impurity scattering in charge or spin channel $\beta (0,x,y,z)$. For a magnetic impurity the scattering channel $\beta$ is fixed by applying a small field $H \ll H_{c2}$ along the $x,y,z$ axis. The conductance channel $\alpha$ is selected by using either a nonmagnetic ($\alpha=0$) or a half-metallic (fully spin polarized) tunneling tip with moment polarized along $\alpha =x,y,z$ and an exchange splitting larger than the heavy fermion quasiparticle band width. Such configuration would allow in principle to determine all elements of the QPI differential conductance tensor. It is given by \cite{byers:93}
\bea
\frac{d\delta I_\alpha(\br,V)}{dV}
&\sim&
 -\frac{1}{\pi}{\rm Im}{\Bigg [}{\rm Tr}_\sigma
 {\Big [}\hrho_\alpha\delta \hat{G}_\beta(\br,\br,\omega = V){\Big ]}_{11}
 {\Bigg ]}
\nonumber\\
&\equiv& \delta N_{\alpha\beta}(\br,\omega). 
\label{eq:conductance}
\eea
Here $\delta \hat{G}_\beta$ is the change of the $4 \times 4$ matrix Green's function in combined Nambu and spin space (each with dimension 2) which is due to impurity scattering in charge or spin channel $\beta (0,x,y,z)$. Furthermore matrix index (11) refers to the Nambu space which results from the trace with respect to $\tau$ including the projector $\frac{1}{2}(1+\tau_z)$. The remaining trace refers to spin space only.

\begin{figure}
  \centering
  \includegraphics[width=0.49\textwidth]{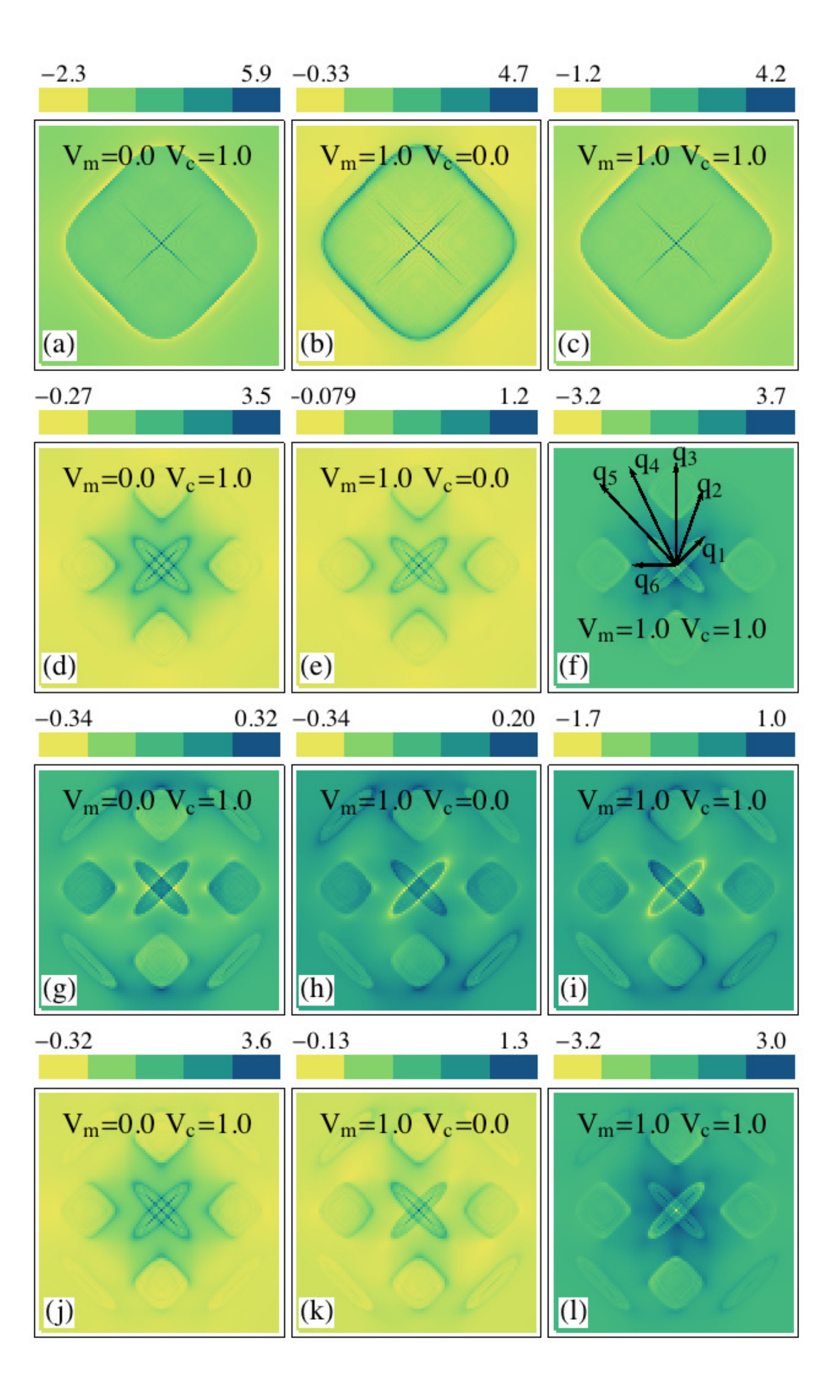}% picture filename
  \caption{ (Color online) 
 Total charge- QPI (${\tilde\Lambda}_{0}(\bq,\omega)$)  for different  scattering processes: first column from non-magnetic impurities, second column from magnetic impurities, and the third row shows mixed non-magnetic/magnetic  impurity scattering.
a-c) for the normal state.
j-l) The same quantity for the superconducting state.
d-f) represent the diagonal contributions  (${\tilde\Lambda}_{0}^d$) and  g-i) represent the off-diagonal contributions   (${\tilde\Lambda}_{0}^a$)  of the total charge- QPI for the superconducting state.
Here $\omega=-0.2t_1$.
 }
\label{Fig5}
\end{figure}

In t-matrix theory the correction to the real-space 
\break
Green's function due to impurity scattering is given by 
\bea
\delta \hG(\br,\om)=\hG_0(\br,\om)\hatt(\om)\hG_0(\br,\om).
\label{eq:greenchange}
\eea
The Fourier transform of differential conductances is then obtained from the QPI functions  ($\bk'=\bk-\bq$)
\bea
\delta N_{\alpha}(\bq,\omega)&=&-\frac{1}{\pi}
 {\rm Im} 
 {\Big [}
 \tL_{\alpha}(\bq,i\omega_n)
 {\Big]}_{i\omega_n\rightarrow \omega + i\delta},
\\
\tL_{\alpha}(\bq,i\omega_n)&=&
\frac{1}{N}\sum_{\bk}
{\rm Tr}_{\s}
{\Big [}
\hrho_\alpha\hat{G}(\bk ,i\omega_n)\hatt(\om)\hat{G}(\bk',i\omega_n)
{\Big ]}_{11},
\nonumber
\label{eq:QPI}
\eea
with $N=L^2$ denoting the number of grid points.  
We assume here that the tunneling happens out of the coherent
heavy quasiparticle states. This is justified for temperature $T$  and frequency $\omega$ much smaller than the Kondo temperature $T^*$ \cite{schmidt:10} which is of the order 14 K for \CP~\cite{bauer:07}. We therefore restrict to frequencies (Figs.~\ref{Fig4}-\ref{Fig6}) of the order of the SC gap and we do not intend to describe the Fano resonance shape \cite{schmidt:10} that appears for higher frequencies  of the order of the effective quasiparticle bandwidth $T^*$.

\subsection{QPI pattern in Born-approximation}
\label{subsec:QPIBorn}

Here we focus on  the spatial oscillations or momentum dependence by weak scattering and ignore the frequency dependence of the t-matrix which has to be included in the strong scattering limit  \cite{liu:08} .
In the Born approximation \cite{akbari:13} we may calculate a general density $\delta N_{\alpha\beta}(\bq,\omega)$ in charge-spin channel $\alpha$ due to impurity scattering in arbitrary but fixed channel $\beta$ given by the frequency independent t-matrix
\bea
\hatt(\bq,\om)=V_\beta(\bq)\hrho_\beta.
\label{eq:tborn}
\eea
Using Eq.~(\ref{eq:QPI}) leads to a density modulation determined by the QPI functions $\tL_\alpha(\bq,\om)= V_\beta(\bq)\Lambda_{\alpha\beta}(\bq,\om)$ according to 
\be
\begin{aligned}
%\nonumber
\delta N_{\alpha\beta}(\bq,\omega)&=-\frac{1}{\pi}V_\beta(\bq)
 {\rm Im} 
 {\Big [}
 \Lambda_{\alpha\beta}(\bq,i\omega_n)
 {\Big]}_{i\omega_n\rightarrow \omega + i\delta},
\\
\Lambda_{\alpha\beta}(\bq,i\omega_n)&=
\frac{1}{N}\sum_{\bk}
{\rm Tr}_{\s}
{\Big [}
\hrho_\alpha\hat{G}(\bk ,i\omega_n)\hrho_\beta\hat{G}(\bk',i\omega_n)
{\Big ]}_{11}.
\label{eq:bornQPI}
\end{aligned}
\ee
%

%%%%%%%%%%%%%%%%%%%%%% figure %%%%%%%%%%%%%%%%%%%%%%%%%%%%%%%%%%%%%%%%%%%%%
%%%%%%%%%%%%%%%%%%%%%%%%%%%%%%%%%%%%%%%%%%%%%%%%%%%%%%%%%%%%%%%%%%%%%%%%

This expression is evaluated by using Eq.~(\ref{eq:greenform}) and performing the remaining $\sigma$- trace we finally get, using the scalar Green's functions in Eq.~(\ref{eq:green}):
%($\bk'=\bk-\bq$)
%
\bea
\label{eq:Lambda}
\Lambda^\bq_{00}(i\omega_n)&=&\frac{1}{4N}\sum_{\bk\xi\xi'}
\bigg[1+\xi\xi'(\hbg_\bk\cdot\hbg_{\bk'})\bigg]K^{\bk\bq}_{\xi\xi'}(i\omega_n),
\nonumber\\
\Lambda^\bq_{{\i}{\i}}(i\omega_n)&=&
\frac{1}{4N}
\sum_{\bk\xi\xi'}
\bigg[1-\xi\xi'(\hbg_\bk\cdot\hbg_{\bk'}-2{\hg}^{\i}_\bk{\hg}^{\i}_{\bk'})\bigg]K^{\bk\bq}_{\xi\xi'}(i\omega_n),
\nonumber\\
\Lambda^\bq_{{\i}0}(i\omega_n)
&=&\frac{1}{2N}\sum_{\bk\xi\xi'}
\xi{\hg}^{\i}_\bk K^{\bk\bq}_{\xi\xi'}(i\omega_n),
\label{eq:QPI2}
\eea
where the integration kernel for intra-band ($\xi=\xi'$) and inter-band ($\xi\neq\xi'$) processes is given by
\be
K^{\bk\bq}_{\xi\xi'}(i\omega_n)=\frac
{(i\omega_n+\e_{\bk\xi})(i\omega_n+\e_{\bk'\xi'})-\Delta_{\bk\xi}\Delta_{\bk'\xi'}}
{[(i\omega_n)^2-E^2_{\bk\xi}] [(i\omega_n)^2-E^2_{\bk'\xi'}]}.
\label{eq:kernel}
\ee

In addition the Rashba term leads to nondiagonal elements in the spin- QPI density, in the tetragonal case with $\bg^z_\bk=0$ to $\Lambda^\bq_{xy}(i\omega_n)=\Lambda^\bq_{yx}(i\omega_n)$ which is evaluated as
\be
\Lambda^\bq_{xy}(i\omega_n)=\frac{1}{4N}\sum_{\bk\xi\xi'}
\xi\xi'({\hg}^x_\bk{\hg}^y_{\bk'}+{\hg}^y_\bk{\hg}^x_{\bk'}) K^{\bk\bq}_{\xi\xi'}(i\omega_n).
\label{eq:Lambdaxy}
\ee
The QPI functions $\Lambda^\bq_{00}$ and  $\Lambda^\bq_{{\i}{\rm j}}$ are even and $\Lambda^\bq_{{\i}0}$ is odd in \bq. The latter needs some special consideration. The real space density  corresponding to Eq.~(\ref{eq:QPI2}) is given by 
\be
\delta N_{{\i}0}(\br,\om)=-\frac{1}{\pi}\int V_c(\bq)\Lambda_{{\i}0}^{(2)}(\bq,\om)\exp(i\bq\br)d\bq,
\label{eq:realdens}
\ee
where we defined $\Lambda_{{\i}0}^{(2)}(\bq,\om)={\rm Im}[ \Lambda_{{\i}0}(\bq,\om)]$ and the non-magnetic scattering potential $V_c(\bq)=V_{cg}(\bq)+V_{cu}(\bq)$ is decomposed into the even (g) part with $V_{cg}(-\bq)=V_{cg}(\bq)$ and the odd part (u) with $V_{cu}(-\bq)=-V_{cu}(\bq)$. 
Since $ \Lambda_{{\i}0}(\bq,\om)$  is odd in \bq~ it is clear that a nonzero spin density in the cross-QPI channel can only arise if the impurity scattering potential has an odd, e.g. p-wave contribution. In this case it is given by
\be
\delta N_{{\i}0}(\br,\om)=-\frac{1}{\pi}\int V_{cu}(\bq)\Lambda_{{\i}0}^{(2)}(\bq,\om)\cos(\bq\br)d\bq.
\label{eq:realdens1}
\ee
Likewise the even part $V_{cg}(\bq)$ leads to the finite real space density contribution for the charge- ($\Lambda^\bq_{00}$) or spin-  ($\Lambda^\bq_{{\i}{\rm j }}$)  QPI functions. In particular for constant $V_c(\bq)=V_c$  no real space spin density modulation can appear from non-magnetic scattering.
We mention that the complementary cross- QPI case, an equivalent charge pattern induced by pure magnetic scattering and described by $\Lambda^\bq_{0{\i}}=\Lambda^\bq_{{\i}0}$ is also possible when $V_m(\bq)$ contains odd contributions. 
From this analysis we expect that QPI for non-centrosymmetric superconductors derived here exhibits a wealth of new effects due to the inversion symmetry breaking Rashba term.\\

\subsection{QPI spectrum with full t-matrix theory}
\label{subsec:QPItmatrix}

For strong scattering the t-matrix becomes frequency dependent and even
resonances may form at impurity sites \cite{akbari:10} which cannot be described within  Born approximation. 
It is also important to ask whether the \bq- space pattern is strongly dependent on the absolute scattering strength and relative
strength of $V_c$ and $V_m$  because this influences the usefulness of QPI for the investigation of the gap function.
Such questions cannot be answered within Born approximation and therefore we now resort to the full t-matrix treatment for QPI functions given in Eq.~(\ref{eq:QPI}). In the special case that the impurity scattering contains only terms due to nonmagnetic  and the z-component of exchange scattering the expressions for the QPI which are assumed \bq-independent according to
\bea
\hV=V_c\hrho_0+V_m\hrho_z.
\label{eq:vmat}
\eea
We will only calculate the charge QPI  function $\tL_0(\bq,\om)$ given in Eq.~(\ref{eq:QPI}) using the full t-matrix of Eq.~(\ref{eq:tmat1}). From the evaluation of matrix products and trace in Nambu space we obtain
\begin{widetext}
\bea
\tL_0(\bq,\om)=\frac{1}{N}\sum_\bk 
&&
{\rm Tr}_\sigma 
\Bigg[
\bigg(t_d^*(\om)G_\bk G_{\bk-\bq}-t_d(\om)F_\bk F_{\bk-\bq}\bigg)
+
\bigg(\tilt_d^*(\om)G_\bk \sigma_z G_{\bk-\bq}-\tilt_d(\om)F_\bk \sigma_z F_{\bk-\bq}\bigg) \nonumber\\
&&-\tilt_a(\om)\bigg(G_\bk \sigma_x F_{\bk-\bq} +F_\bk \sigma_x G_{\bk-\bq} \bigg)
-t_a(\om)\bigg(G_\bk (i\sigma_y) F_{\bk-\bq} +F_\bk (i\sigma_y) G_{\bk-\bq} \bigg)
\Bigg].
\label{eq:tQPI}
\eea
In the case of purely nonmagnetic scattering ($V_m=0$)  $\tilt_d(i\omega_n)=\tilt_a(i\omega_n)=0$ and the expression reduces
to
\be
\tL_0(\bq,\om)=\frac{1}{N}\sum_\bk 
{\rm Tr}_\sigma \Bigg[
\bigg(t_d^*(\om)G_\bk G_{\bk-\bq}-t_d(\om)F_\bk F_{\bk-\bq}\bigg)
-t_a(\om)\bigg(G_\bk (i\sigma_y) F_{\bk-\bq} +F_\bk (i\sigma_y) G_{\bk-\bq} \bigg)
\Bigg].
\label{eq:tQPI1}
\ee
%\end{widetext}
%
In the following discussion we restrict to the tetragonal Rashba systems where  $\hbg_\bk\cdot\hbz =0$, i.e. the Rashba vector is perpendicular to the impurity moment $\langle S_z\rangle$ associated with magnetic scattering  $V_m$. In this case terms involving
$\tilt_d(\om)$ do not appear in $\tL_0(\bq,\om)$. The general case with all three components of the Rashba vector $\hbg_\bk$ present is treated in \ref{sec:appA}.
Using the Green's function matrices in Eq.~(\ref{eq:greenform}) and performing the traces in spin space this leads to $(\bk'=\bk-\bq)$
%
%\begin{widetext}
\bea
\hspace{-1cm}
\tL_0(\bq,\om)=
&&
t_d^*(\om)
\frac{1}{N}\sum_\bk
\bigg[G_+^\bk G_+^{\bk'}+(\hbg_\bk\cdot\hbg_\bk')G_-^\bk G_-^{\bk'}\bigg]
+t_d(\om)\frac{1}{N}\sum_\bk\bigg[F_+^\bk F_+^{\bk'}+(\hbg_\bk\cdot\hbg_\bk')F_-^\bk F_-^{\bk'}\bigg] 
\nonumber \\&&
-\frac{2}{N}\sum_{\bk} {\hg}_{\bk}^x
\bigg[{\hg}^x_{\bk'}t_a(\om)+i{\hg}^y_{\bk'}\tilt_a(\om)\bigg]
G_-^{\bk} F_-^{\bk'}.
\label{eq:tQPI2}
\eea
Inserting the explicit expressions for the normal and anomalous Green's functions in Eq.~(\ref{eq:green}) we finally obtain a QPI function that consists of two contributions, 
\bea
\tL_0(\bq,\om)&=&\tL^d_0(\bq,\om)+\tL^a_0(\bq,\om) ,
\label{eq;tQPI3}
\eea
originating in the diagonal (d) and anti-diagonal (a) terms of the t-matrix in Eqs.~(\ref{eq:tmat1}).
\be
\begin{aligned}
\tL^d_{0}(\bq,i\omega_n)&=\frac{1}{4N}\sum_{\bk\xi\xi'}
\bigg[1+\xi\xi'(\hbg_\bk\cdot\hbg_{\bk'})\bigg]
\frac{t_d^*(\om)(i\omega_n+\e_{\bk\xi})(i\omega_n+\e_{\bk'\xi'})-t_d(\om)\Delta_{\bk\xi}\Delta_{\bk'\xi'}}
{[(i\omega_n)^2-E^2_{\bk\xi}] [(i\omega_n)^2-E^2_{\bk'\xi'}]}, 
\\
\tL^a_{0}(\bq,i\omega_n)&=-\frac{1}{2N}\sum_{\bk\xi\xi'}
\xi\xi'{\hg}^x_\bk\bigg[ {\hg}^x_{\bk'}t_a(\om)+i{\hg}^y_{\bk'}\tilt_a(\om)\bigg]
\frac{(i\omega_n+\e_{\bk\xi})\Delta_{\bk'\xi'}}
{[(i\omega_n)^2-E^2_{\bk\xi}] [(i\omega_n)^2-E^2_{\bk'\xi'}]}.
\label{eq:tQPI4}
\end{aligned}
\ee
\end{widetext}
We can check this result for the Born approximation where $t_d(\om)=t^*_d(\om)=V_c$ and   $t_a(\om)=\tilt_a(\om)=0$ (see also \ref{sec:appB}). Then the fraction in the first equation is given by $V_cK^{\bk\bq}_{\xi\xi'}(i\omega_n)$, furthermore 
$\Lambda^a_{0}(\bq,i\omega_n)=0$ and therefore  $\tL_0(\bq,\om)=V_c\Lambda_{00}(\bq,\om)$ reduces to the previous result in
Eq.~(\ref{eq:QPI2}). For nonmagnetic scattering only $(V_m=0)$ the anti-diagonal (a) contribution in Eq.~(\ref{eq:tQPI4}) simplifies because $\tilt_a(\om)=0$. Then we may write
\be
\tL^a_{0}(\bq,i\omega_n)=
\sum_{\bk\xi\xi'}
\frac{-\xi\xi'g^x_\bk g^x_{\bk'}}{2N}
\frac{t_a(\om)(i\omega_n+\e_{\bk\xi})\Delta_{\bk'\xi'}}
{[(i\omega_n)^2-E^2_{\bk\xi}] [(i\omega_n)^2-E^2_{\bk'\xi'}]}.
\label{eq:tQPI5}
\ee
Note that the even in the general case the anti-diagonal QPI contribution $\Lambda^a_{0}(\bq,i\omega_n)$ is only nonzero in the superconducting phase because it is due to Andreev-type scattering processes that require the presence of a condensate. 
Since the superconducting gap has nontrivial $A_{2u}$ symmetry one has to expect that $\tL^a_{0}(\bq,i\omega_n)$ is not fully symmetric.

In the strong scattering case bound states or resonances in the superconducting gap can appear which may be investigated by the calculation of the change of the local density of states  (LDOS), $\delta N_0(\omega)$, at the impurity site. It is given by the integral over the real space density oscillations or the momentum space integral over the corresponding charge- QPI function according to
\be
\delta N_{0}(\omega)=-\frac{1}{\pi}{\rm Im} 
{\Big [}
\frac{1}{N}\sum_\bq\tL_0(\bq,\om)
{\Big ]}
.
\label{eq:LDOS}
\ee
The total LDOS is then given by $N_{0t}(\omega)=N_0(\omega)+\delta N_0(\omega)$ where $N_0(\omega)=-(1/\pi){\rm Im} [g(\om)]_{i\omega_n\rightarrow \omega + i\delta}$ is the background DOS (per site) of quasiparticle states.

%%%%%%%%%%%%%%%%%%%%%% figure %%%%%%%%%%%%%%%%%%%%%%%%%%%%%%%%%%%%%%%%%%%%%
\begin{figure}
  \centering
  \includegraphics[width=0.5\textwidth]{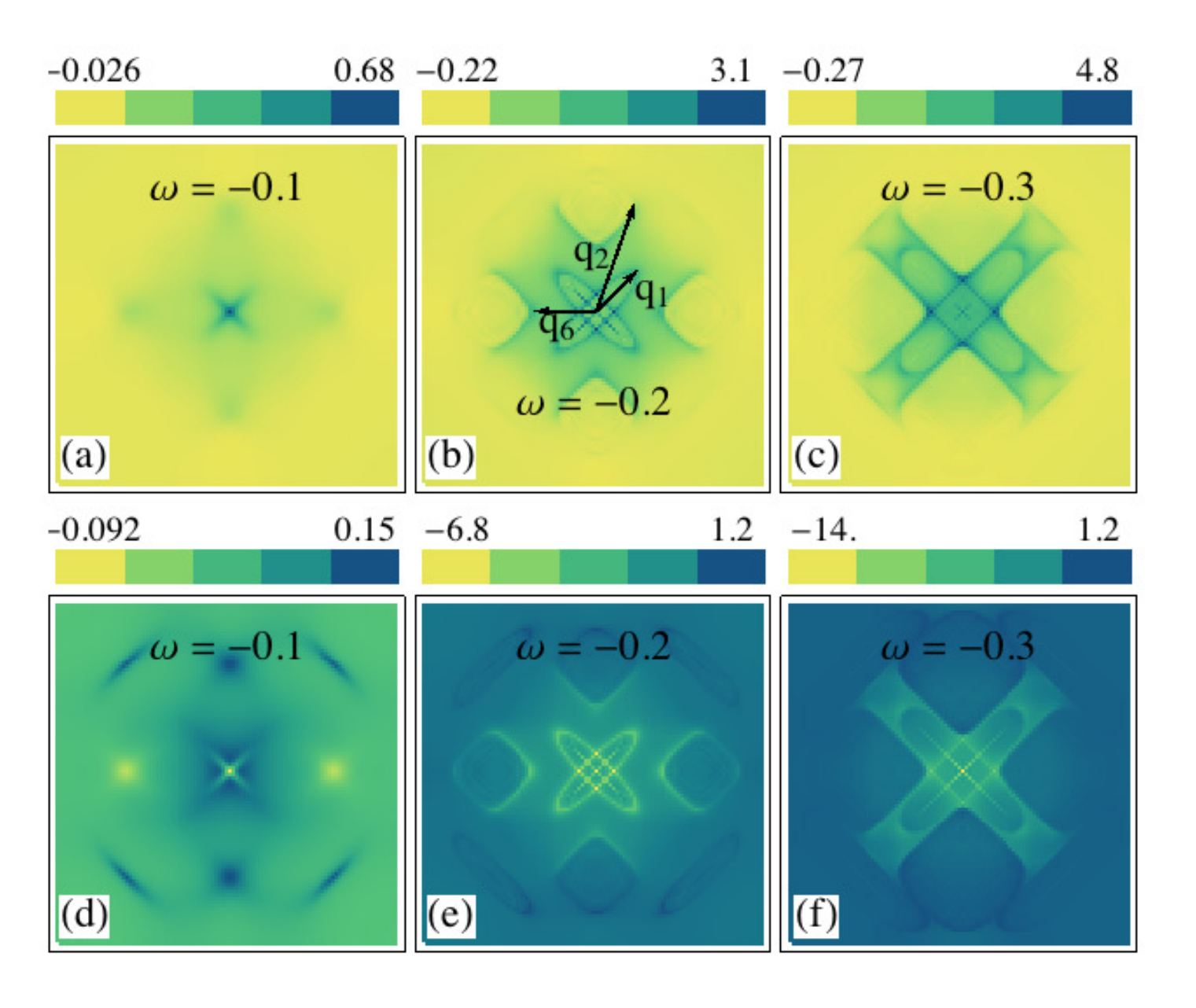}% picture filename
  \caption{(Color online) 
  Comparison of weak (top) and strong (bottom) scattering cases.
  a-c) Total charge- QPI (${\tilde\Lambda}_{0}(\bq,\omega)$) in the superconducting  state for different $\omega<0$ and  scattering from non-magnetic impurities in Born approximation.
d-f)  Total charge- QPI (${\tilde\Lambda}_{0}(\bq,\omega)$) in the superconducting  state for different $\omega<0$ and  scattering from non-magnetic impurities using t-matrix formalism with  $V_c=5t_1$ and $V_m=0$.
\vspace{0.1cm}
}
\label{Fig6}
\end{figure}

%%%%%%%%%%%%%%%%%%%%%%fig%%%%%%%%%%%%%%%%%%%%%%%%%%%%%%%%%%%%%%%%%%%%%%%%

%%%%%%%%%%%%%%%%%%%%%% figure %%%%%%%%%%%%%%%%%%%%%%%%%%%%%%%%%%%%%%%%%%%%%
\begin{SCfigure*}
  \centering
  \includegraphics[width=0.70\textwidth]{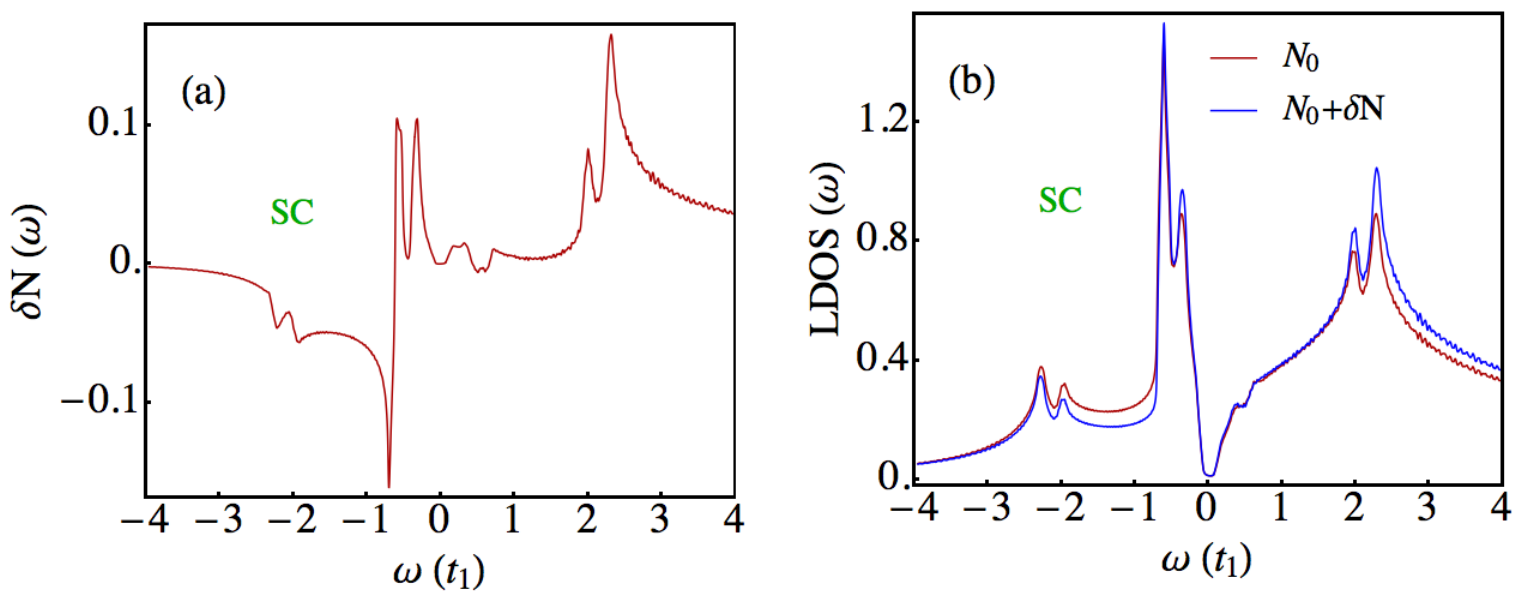}% picture filename
  \caption{(Color online) 
  LDOS in the superconducting state for $V_c=t_1$ a) correction due to impurity scattering (c.f. Fig.~\ref{Fig3}b). b) Background DOS $N_0$ and total LDOS.
\vspace{2cm}
}
\label{Fig7}
\end{SCfigure*}

%%%%%%%%%%%%%%%%%%%%%%fig%%%%%%%%%%%%%%%%%%%%%%%%%%%%%%%%%%%%%%%%%%%%%%%%

\section{Numerical results and discussion}
\label{sec:numerical}

In the presentation of numerical results we focus on the charge-QPI $\tL_0(\bq,\om)$ which is in any case experimentally the most easily accessible quantity. We  discuss primarily the results obtained by using the full t-matrix theory for which we have derived a closed and explicit representation previously. We emhasize that we do not give an exhaustive discussion here of all possible situations that may occur on changing the electronic ($t_1, t_2, g$), superconducting ($\psi_0,\psi_1,\phi_0$) and impurity scattering parameters ($V_c,V_m$) and also the polarization channels $\alpha,\beta$. We rather fix the first two sets of parameters to obtain a realistic model of \CP~ and then study a few typical QPI spectra as function of frequency (bias voltage) $\omega$ and scattering parameters ($V_c,V_m$). We will mostly investigate the nonmagnetic scattering $V_c$ only and restrict to the scalar charge QPI function, i.e., we focus on the $(\alpha,\beta)=(0,0)$ channel of Eq.~(\ref{eq:conductance}).
The salient features of the additional spin- QPI and cross-QPI have been discussed previously within Born approximation \cite{akbari:13}. They are experimentally more difficult to access because they require the spin polarization of both the impurity spin, e.g. by a small external field as well as a spin analyzer for the tunneling current.%\\

The underlying model is summarized by the panels of Fig.~\ref{Fig1} which show the Rashba split bands (a) and Fermi surface (b) in the normal state. The model parameters are choosen \cite{takimoto:08,takimoto:09} such that the M-point Fermi surface of \CP~ is reproduced qualitatively.  The dashed lines show the nodes of the gap functions $\Delta_{\bk\xi}$. The gap parameters are tuned such that nodes for $\Delta_{\bk -}$ appear on the $\epsilon_{\bk-}$ FS sheet. The existence of nodes is suggested e.g. by thermal conductivity measurements \cite{izawa:05}. The surfaces of constant quasiparticle energies in the superconducting phase (c) consist of thin sheets with points of large curvature in \bk-space. The connecting $\bq_i$ vectors of these extremal points are the ones which will show up prominently in the QPI spectra.%\\

A basic ingredient of the QPI theory are the momentum integrated Green's functions given in Eq.~(\ref{eq:gintmat}).
The energy dependence of $f$ and $g$ is shown in Fig.~\ref{Fig2}.  In Fig.~\ref{Fig2}a the real and imaginary part of $-g(\om)$ in the normal state are plotted where the unperturbed density of states (DOS) of Rashba bands is given by 
$$N_0(\omega)=-\frac{1}{\pi}{\rm Im} {\Big [}g(\om){\Big ]}_{i\omega_n\rightarrow \omega + i\delta}=\frac{1}{N}\sum_{\bk\xi}\delta(\omega-\epsilon_{\bk\xi}).$$ 
 This indicates that all calculations, also for QPI spectrum, are done for zero temperature by analytic continuation to the real axis. Numerically we use a finite imagninary part $\delta =0.005t_1$.
The DOS has two peaks at $\omega\simeq\pm t_1$ which are due to the van Hove singularities at X and M points in Fig.~\ref{Fig1}a. In the superconducting state the DOS ( Fig.~\ref{Fig2}b) changes considerably due to the gap opening of the order $|\Delta_{\bk\xi}|\simeq t_1$. 
Due to the large gap and the strong particle-hole asymmetry of the dispersion around the Fermi level the nodal V-shape of the DOS is asymmetric around $\omega=0$ and exists only for small frequency $\omega\ll \Delta_{\bk-}$. The corresponding anomalous Green's function $-f(\omega)$ is shown in  Fig.~\ref{Fig2}c. 

From the momentum integrated Green's functions the t-matrix components of Eq.~(\ref{eq:telement})  that determine the QPI functions can directly be obainted once the scattering model is fixed by $(V_c,V_m)$ parameters.
The element $t_d(\omega)$ which enters in the diagonal part of the QPI function is shown in  Fig.~\ref{Fig2} for the normal (a) and superconducting state (b). Its imaginary part still resembles the DOS with a cutoff at the lower M-point band edge and the peak due to the X-point singularity.%\\

In Fig.~\ref{Fig4} we show the QPI spectrum for nonmagnetic ($V_c$) scattering and increasingly negative frequencies or voltages (from left to right).   The top and bottom rows show the total charge-QPI function $\tL_0(\bq,\om)$ in the normal and superconducting state, respectively. The latter has two contributions, the diagonal $\tL^d_0(\bq,\om)$ shown in the second row and the anti-diagonal part  $\tL^a_0(\bq,\om)$ presented in the third row (it vanishes identically in the normal state according to   Eq.~(\ref{eq:tQPI4})).
The first row basically shows the $``2k_F"$ contour resulting from scattering across the M-point constant energy surfaces in Fig.~\ref{Fig1}b. The dimensions of these contours decrease with increasingly negative frequency when the bottom of the band at the M-point is approached.
In addition one can clearly see a diagonal cross feature that results from small wave vector intra band ($\xi=\xi'$) scattering parallel to the M-point sheet.

In the superconducting state (second row) the QPI looks very different due to the breakup of constant quasiparticle surfaces caused by the gap opening (see Fig.~\ref{Fig1}c). The maximum amplitude of the spectrum is dominated by the wave vectors $\bq_i$ that correspond to connections between points of maximum curvature on the broken sheets as indicated in  Fig.~\ref{Fig4}e. Therefore the QPI spectrum is a kind of map of the reconstructed surfaces of constant quasiparticle energies and highly specific for the node structure of the gap function. In particular the wave vector $\bq_1$ corresponds to scattering connecting opposite sites of the accidental gap node. The observation of such QPI feature would be a clear evidence for the existence and position of the node points (lines in 3D which are sofar only conjectures from low temperature transport measurements).

In the third row we show the anti-diagonal contribution in the superconducting state which has similar overall features as the second row. One additional aspect is that it exhibits directly the spatial (reflection) symmetry breaking of the triplet $A_{2u}$ state. However the anti-diagonal  part has considerably lower amplitude than the diagonal one so that the total QPI function   $\tL_0(\bq,\om)$ corresponding to the experimental conductance is dominated by the diagonal part. It is shown in the last row of   Fig.~\ref{Fig4} for the superconducting state.\\

The frequency dependence of QPI pattern in Fig.~\ref{Fig4} is only shown for the nonmagnetic ($V_c$) scattering. In general the scattering potentials enter in the combinations $V_c\pm V_m$ into the t-matrix elements of  Eq.~(\ref{eq:telement}). Therefore we now fix the frequency to $\omega=-0.2t_1$ and investigate the QPI spectrum as function of various scattering strengths $(V_c,V_m)$. We compare purely nonmagnetic  ($V_c$) scattering as before, equal scattering strength $V_c=V_m$ and purely magnetic ($V_m$) scattering in Fig.~\ref{Fig5} ($V_c,V_m$ are given by $t_1$ units).  We show again the total charge-QPI function  $\tL_0(\bq,\om)$ of the normal and superconducting state in the top and bottom row, respectively. The latter has diagonal and anti-diagonal contributions which are shown in the second and third row, respectively. 
 We observe that the amplitude of the QPI functions vary considerably with ($V_c,V_m)$ combination, however the momentum {\it pattern} stays surprisingly similar in the three cases, determined by the shape of split Fermi surface sheets and the nodal gap structure. This supports the view that also in the non-centrosymmetric superconductors the QPI spectrum may be used to investigate the SC gap structure and is not veiled by the influence of the details of scattering mechanism.\\

In addition to the composition of the scattering potential it is important to understand the effect of its overall strength. Most QPI investigations use only the Born approximation corresponding to weak scattering potentials. On the other hand for strong scattering the integrated QPI spectra or LDOS may exhibit resonance peaks as function of frequency under suitable conditions. It is therefore important to know to which extent the momentum and frequency pattern of  $\tL_0(\bq,\om)$ depends on the strength of the scattering potential and the type of approximation. In Fig.~\ref{Fig6} we show the comparison of weak scattering in Born approximation (top row) and strong scattering in full t-matrix theory (bottom row) for the SC state. From the comparison we conclude that for all frequencies the \bq -space pattern of QPI function are remarkably similar although the amplitudes are reversed (Fig.~\ref{Fig6}d-f is the `negative' of Fig.~\ref{Fig6}a-c). This may be understood from the limit of $t_d(\om)^*$ in Eq.~(\ref{eq:tQPI4}) which is $V_c>0$ in Born approximation and $-1/g(\om)$ in the strong scattering t-matrix approximation. In the frequency range shown in  Fig.~\ref{Fig6} this quantity is $<0$ according to Fig.~\ref{Fig3}b. 
Our results of  Figs.~\ref{Fig5},\ref{Fig6}   suggests that the observed QPI pattern is rather insensitive to the details of the scattering mechanism and is mainly determined by Fermi surface structure and superconducting gap nodes.
  Fig.\ref{Fig6} thereby represents the extreme case of very strong scattering. However this conclusion is also valid for cases of intermediate scattering strength.
Therefore we believe that also in the case of noncentrosymmetric superconductors the STM-QPI technique may be used to investigate the symmetry of the superconducting gap.\\

Finally we also show the total LDOS as function of energy (Fig.~\ref{Fig7}) which is obtained from the  integrated QPI spectrum according to Eq.~(\ref{eq:LDOS}). The change $\delta N({\omega})$ due to normal impurity scattering is shown in  Fig.~\ref{Fig7}a and the total LDOS $N(\omega)$ in comparison to the unperturbed quasiparticle DOS $N_0(\omega)$ is presented in  Fig.~\ref{Fig7}b for the superconducting phase. The changes are relatively small and in particular do not show the development of a separate resonance peak for the parameter range investigated.
This may be due to the nodal structure of the gap which leads to considerable imaginary part in $t_d(\omega)$ even at small frequencies (Fig.~\ref{Fig3}b).

\section{Conclusion and outlook}
\label{sec:conclusion}

In this work we have derived the full t-matrix theory  of quasiparticle interference in non-centrosymmetric superconductors.   Unlike in common, purely numerical treatments  we have succeeded to give a closed analytical representation for the full t-matrix that shows explicitly how the combination of normal and magnetic scattering determines the QPI spectrum, and in particular how the non-diagonal Andreev scattering terms in QPI appear beyond Born approximation as an effect of the condensate.
For our closed expression for the charge-QPI function only one remaining momentum space integration to obtain the integrated normal and anomalous Green's functions needs to be performed numerically. Our theory is valid for an arbitrary combination and strength of non-magnetic and magnetic scattering potentials where the impurity spin is polarized perpendicular to the Rashba-vector $\bg_\bk$. Sofar the non-centrosymmetric QPI problem has only been treated within Born approximation \cite{akbari:13} where the frequency dependence of the scattering is neglected. It therefore remained an open problem whether this influences the interpretation of the momentum dependence of QPI pattern. Our results show the the latter is remarkably stable and qualitatively unchanged by frequency variation and modification of the combination of scattering potentials $(V_c,V_m)$. \\

Furthemore the QPI structure pattern and characteristic wave vectors connected with gap features are almost identical in the Born approximation and full t-matrix theory. Consequently for practical purposes one may assume that the frequency dependence and momentum dependence of QPI pattern shows no strong interdependence and this holds true for arbitrary scattering strengths.
Therefore our theory demonstrates that QPI may be used as a stable method in non-centrosymmetric superconductors to investigate the symmetry of the superconducting gap function and its accidental node positions   which cannot be determined by the other applicable methods.\\

 We have discussed the results of our analytical  t-matrix and QPI spectrum theory for typical situations without giving a full systematic survey of the predictions as function of model (scattering potential, gap function) parameters. For that program to be carried out in a sensible way first experimental results for NCS superconductors  are needed for orientation. As a starting point one should consider the normal (and nonmagnetic) state and see whether the predictions in Fig.\ref{Fig4}a-c can be verified, i.e. whether the simple 2D FS model used here is a reasonable simplification. In a next step it is necessary to verify that the influence of small-moment magnetic order (which has been neglected here) is unimportant. And only as final step one may hope to gain insight into the question of accidental node extistence and their postions by employing a more extensive search in the parameter space of the the model analyzed here.

We finally mention that our theory is not restricted to the tetragonal Ce-based 131 compounds but may straightforwardly be extended to cubic non-centrosymmetric compounds \cite{yuan:06} like Li$_2$Pd$_3$B and Li$_2$Pt$_3$B   with general Rashba vector as discussed in \ref{sec:appA}.

%\appendix
\section*{Appendix A}
\label{sec:appA}

Here we give the complete expression for the charge-QPI function in the full t-matrix case. Contrary to Sec.~\ref{subsec:QPItmatrix} where we treated the tetragonal symmetry case with $g_\bk^z =0$  we now do not pose any condition on the Rashba vector. In the general case we have to add the contributions coming from nonzero $g_\bk^z$ to Eq.~(\ref{eq:tQPI4}). They are present, e.g., in the cubic non-centrosymmetric superconductors and they are given by $(\bk'=\bk-\bq)$
\begin{widetext}
\be
\tL^z_0(\bq,\om)=\frac{2}{N}\sum_\bk{\hg}^z_\bk
\bigg[\tilt_d^*(\om)G_-^\bk G_+^{\bk'}
-\tilt_d(\om)F_+^\bk F_+^{\bk'}\bigg]
-\frac{2}{N}\sum_{\bk} {\hg}_{\bk}^z
\bigg[\tilt_a(\om)+{\hg}^z_{\bk'}t_a(\om)\bigg]
G_-^{\bk} F_-^{\bk'}.
\label{eq;appA1}
\ee
The first (diagonal d) and second (anti-diagonal a) terms may be evaluated explicitly as
\be
\begin{aligned}
\hspace{-1.3cm}
\tL^{zd}_{0}(\bq,i\omega_n)&=\frac{1}{2N}\sum_{\bk\xi\xi'}
\xi{\hg}^z_\bk
\frac{\tilt_d^*(\om)(i\omega_n+\e_{\bk\xi})(i\omega_n+\e_{\bk'\xi'})-\tilt_d(\om)\Delta_{\bk\xi}\Delta_{\bk'\xi'}}
{[(i\omega_n)^2-E^2_{\bk\xi}] [(i\omega_n)^2-E^2_{\bk'\xi'}]}, 
\\
\hspace{-1.3cm}
\tL^{za}_{0}(\bq,i\omega_n)&=-\frac{1}{2N}\sum_{\bk\xi\xi'}
\xi\xi'{\hg}^z_{\bk'}\bigg[\tilt_a(\om)+{\hg}^z_{\bk}t_a(\om)\bigg]
\frac{(i\omega_n+\e_{\bk\xi})\Delta_{\bk'\xi'}}
{[(i\omega_n)^2-E^2_{\bk\xi}] [(i\omega_n)^2-E^2_{\bk'\xi'}]}.
\label{eq:appA2}
\end{aligned}
\ee
The total QPI function contributions in the general case of arbitrary $\bg_\bk$ are then given by the sum of Eqs.~(\ref{eq:tQPI4},\ref{eq:appA2}), namely $\tL^{d}_{0t}(\bq,i\omega_n)=\tL^{d}_{0}(\bq,i\omega_n)+\tL^{zd}_{0}(\bq,i\omega_n)$ and
 $\tL^{a}_{0t}(\bq,i\omega_n)=\tL^{a}_{0}(\bq,i\omega_n)+\tL^{za}_{0}(\bq,i\omega_n)$. The meaning of the first term becomes clear when we consider the Born approximation. In this case $\tilt_d(\om)=V_m$ and we get $\tL^{zd}_{0}(\bq,i\omega_n)=V_m\Lambda^\bq_{0z}(\om)$ which corresponds to the cross-QPI function of Eq.~(\ref{eq:QPI2}) describing the charge modulation due to exchange scattering $V_m$ from impurity spins $S_z$. Note that this is equal to the inverse process (spin modulation from non-magnetic scattering $V_c$), i.e. $\Lambda^\bq_{0z}(\om)=\Lambda^\bq_{z0}(\om)$. In Born approximation the anti-diagonal contribution $\tL^{za}_{0}(\bq,\om)$ vanishes. \\
 
 In the full t- matrix theory, but assuming only nonmagnetic scattering $(V_m=0)$ this term simplifies and adding it to the tetragonal 
terms of Eq.~(\ref{eq:tQPI4})  we get an expression similar to  Eq.~(\ref{eq:tQPI5})
\bea
\hspace{-1cm}
\tL^a_{0t}(\bq,i\omega_n)&=&-\frac{1}{2N}\sum_{\bk\xi\xi'}
\xi\xi'(g^x_\bk g^x_{\bk'}+g^z_\bk g^z_{\bk'})
\frac{t_a(\om)(i\omega_n+\e_{\bk\xi})\Delta_{\bk'\xi'}}
{[(i\omega_n)^2-E^2_{\bk\xi}] [(i\omega_n)^2-E^2_{\bk'\xi'}]}.
\label{eq:appA3}
\eea
%\end{widetext}

\section*{Appendix B}
\label{sec:appB}
Here we give a different form of the full t-matrix QPI functions in Eq.~(\ref{eq:tQPI4})  that makes its connection to the result from Born approximation more transparent. Using $t_d(\om)=t_d(\om)'+it_d(\om)''$ we can write the imaginary part of the diagonal term as
%\begin{widetext}
\be
\hspace{-.0cm}
{\rm Im}{\Big [}\tL^d_{0}(\bq,i\omega_n){\Big ]}=\frac{1}{4N}\sum_{\bk\xi\xi'}
\bigg(1+\xi\xi'(\hbg_\bk\cdot\hbg_{\bk'})\bigg)
\bigg[t_d(\om)' {\rm Im}{\Big [}K^{\bk\bq}_{\xi\xi'}(\om)_-{\Big ]}+t_d(\om)'' {\rm Re}{\Big [}K^{\bk\bq}_{\xi\xi'}(\om)_+{\Big ]}\bigg],
\ee
\end{widetext}
where we have now two integration kernels defined by
\be
K^{\bk\bq}_{\xi\xi'}(i\omega_n)_\pm=\frac
{(i\omega_n+\e_{\bk\xi})(i\omega_n+\e_{\bk'\xi'})\pm\Delta_{\bk\xi}\Delta_{\bk'\xi'}}
{[(i\omega_n)^2-E^2_{\bk\xi}] [(i\omega_n)^2-E^2_{\bk'\xi'}]}.
\label{eq:appB1}
\ee
In Born approximation the scattering reduces to $t_d(\om)'=V_c$ and $t_d(\om)''=0$. Furthermore  $t_a(\om)\equiv 0$. Then we have only the diagonal part and $\tL_0(\bq,\om)=V_c\Lambda_{00}^\bq(\om)$ where $\Lambda_{00}^\bq(\om)$ is now given again by
Eqs.(\ref{eq:QPI2},\ref{eq:kernel}) with the identification $K^{\bk\bq}_{\xi\xi'}(i\omega_n)\equiv K^{\bk\bq}_{\xi\xi'}(i\omega_n)_-$. Therefore the $K_+$ kernel in the QPI function can only appear beyond Born approximation.

%
% BibTeX users please use

\bibliographystyle{Ref}
\bibliography{References}

\begin{thebibliography}{43}

\bibitem{damascelli:03}
A.~Damascelli, Z.~Hussain, Z.X. Shen, Rev. Mod. Phys. \textbf{75}, 473 (2003)

\bibitem{okazaki:12}
K.~Okazaki, Y.~Ota, Y.~Kotani, W.~Malaeb, Y.~Ishida, T.~Shimojima, T.~Kiss,
  S.~Watanabe, C.T. Chen, K.~Kihou et~al., Science \textbf{337}, 1314 (2012)

\bibitem{matsuda:06}
Y.~Matsuda, K.~Izawa, I.~Vekhter, J. Phys. Condens. Matter \textbf{18}, R705
  (2006)

\bibitem{McElroy:2003}
K.~McElroy, R.W. Simmonds, J.E. Hoffman, D.H. Lee, J.~Orenstein, H.~Eisaki,
  S.~Uchida, J.C. Davis, Nature (London) \textbf{422}, 592 (2003)

\bibitem{Hanaguri:2009}
T.~Hanaguri, Y.~Kohsaka, M.~Ono, M.~Maltseva, P.~Coleman, I.~Yamada, M.~Azuma,
  M.~Takano, K.~Ohishi, H.~Takagi, Science \textbf{323}, 923 (2009)

\bibitem{Hanaguri:2010}
T.~Hanaguri, S.~Niitaka, K.~Kuroki, H.~Takagi, Science \textbf{328}, 474 (2010)

\bibitem{allan:12}
M.P. Allan, A.W. Rost, A.P. Mackenzie, Y.~Xie, J.C. Davis, K.~Kihou, C.H. Lee,
  A.~Iyo, H.~Eisaki, T.M. Chuang, Science \textbf{336}, 563 (2012)

\bibitem{byers:93}
J.M. Byers, M.E. Flatt\'e, D.J. Scalapino, Phys. Rev. Lett. \textbf{71}, 3363
  (1993)

\bibitem{capriotti:03}
L.~Capriotti, D.J. Scalapino, R.D. Sedgewick, Phys. Rev. B \textbf{68}, 014508
  (2003)

\bibitem{pereg:03}
T.~Pereg-Barnea, M.~Franz, Phys. Rev. B \textbf{68}, 180506 (2003)

\bibitem{Wang:2003}
Q.H. Wang, D.H. Lee, Phys. Rev. B \textbf{67}, 020511 (2003)

\bibitem{zhu:04}
L.~Zhu, W.A. Atkinson, P.J. Hirschfeld, Phys. Rev. B \textbf{69}, 060503(R)
  (2004)

\bibitem{nunner:06}
T.S. Nunner, W.~Chen, B.M. Andersen, A.~Melikyan, P.J. Hirschfeld, Phys. Rev. B
  \textbf{73}, 104511 (2006)

\bibitem{Maltseva:2009}
M.~Maltseva, P.~Coleman, Phys. Rev. B \textbf{80}, 144514 (2009)

\bibitem{andersen:09}
B.M. Andersen, P.J. Hirschfeld, Phys. Rev. B \textbf{79}, 144515 (2009)

\bibitem{Akbari:2010}
A.~Akbari, J.~Knolle, I.~Eremin, R.~Moessner, Phys. Rev. B \textbf{82}, 224506
  (2010)

\bibitem{Knolle:2010}
J.~Knolle, I.~Eremin, A.~Akbari, R.~Moessner, Phys. Rev. Lett. \textbf{104},
  257001 (2010)

\bibitem{huang:11}
H.~Huang, Y.~Gao, D.~Zhang, , C.S. Ting, Phys. Rev. B \textbf{84}, 134507
  (2011)

\bibitem{akbari:11}
A.~Akbari, P.~Thalmeier, I.~Eremin, Phys. Rev. B \textbf{84}, 134505 (2011)

\bibitem{allan:13}
M.P. Allan, F.~Massee, D.K. Morr, J.~van Dyke, A.~Rost, A.P. Mackenzie,
  C.~Petrovic, J.C. Davis, Nature Physics \textbf{9}, 468 (2013)

\bibitem{zhou:13}
B.B. Zhou, S.~Misra, E.H. da~Silva~Neto, P.~Aynajian, R.E. Baumbach, J.D.
  Thompson, E.D. Bauer, A.~Yazdani, Nature Physcis \textbf{9}, 474 (2013)

\bibitem{aynajian:10}
P.~Aynajian, E.H. da~Silva~Neto, C.V. Parker, Y.K. Huang, A.~Pasupathy, J.A.
  Mydosh, A.~Yazdani, Proc. Natl. Acad. USA \textbf{107}, 10383 (2010)

\bibitem{schmidt:10}
A.R. Schmidt, M.H. Hamidian, P.~Wahl, F.~Meier, A.V. Balatsky, J.D. Garrett,
  T.J. Williams, G.M. Luke, J.C. Davis, Nature \textbf{465}, 570 (2010)

\bibitem{yuan:12}
T.~Yuan, J.~Figgins, D.K. Morr, Phys. Rev. B \textbf{86}, 035129 (2012)

\bibitem{settai:07}
R.~Settai, T.~Takeuchi, Y.~Onuki, J. Phys. Soc. Jpn \textbf{76}, 051003 (2007)

\bibitem{bauer:07}
E.~Bauer, H.~Kaldarar, A.~Prokofiev, E.~Royanian, A.~Amato, J.~Sereni,
  W.~Br\"amer-Escamilla, I.~Bonalde, J. Phys. Soc. Jpn \textbf{76}, 051009
  (2007)

\bibitem{kimura:07}
N.~Kimura, Y.~Muro, H.~Aoki, J. Phys. Soc. Jpn \textbf{76}, 051010 (2007)

\bibitem{akbari:13}
A.~Akbari, P.~Thalmeier, Europhysics Letters \textbf{102}, 57008 (2013)

\bibitem{frigeri:06}
P.A. Frigeri, D.F. Agterberg, I.~Milat, M.~Sigrist, Eur. Phys. J. B
  \textbf{54}, 435 (2006)

\bibitem{eremin:06}
I.~Eremin, J.F. Annett, Phys. Rev. B \textbf{74}, 184524 (2006)

\bibitem{fujimoto:07}
S.~Fujimoto, J. Phys. Soc. Jpn \textbf{76}, 051008 (2007)

\bibitem{samokhin:04}
K.V. Samokhin, E.S. Zijlstra, S.K. Bose, Phys. Rev. B \textbf{69}, 094514
  (2004)

\bibitem{sigrist:09}
M.~Sigrist, \emph{Introduction to unconventional superconductivity in
  non-centrosymmetric superconductors} (American Institute of Physics, 2009),
  chap. CP1162, pp. 55--97

\bibitem{takimoto:08}
T.~Takimoto, J. Phys. Soc. Jpn \textbf{77}, 113706 (2008)

\bibitem{takimoto:09}
T.~Takimoto, P.~Thalmeier, J. Phys. Soc. Jpn \textbf{78}, 103703 (2009)

\bibitem{yanase:08}
Y.~Yanase, M.~Sigrist, J. Phys. Soc. Jpn \textbf{77}, 124711 (2008)

\bibitem{izawa:05}
K.~Izawa, Y.~Kasahara, Y.~Matsuda, K.~Behnia, T.~Yasuda, R.~Settai, Y.~Onuki,
  Phys. Rev. Lett. \textbf{94}, 197002 (2005)

\bibitem{sigrist:96}
M.~Sigrist, M.E. Zhitomirsky, J. Phys. Soc. Jpn \textbf{65}, 3452 (1996)

\bibitem{mineev:99}
V.P. Mineev, K.V. Samokhin, \emph{Introduction to unconventional
  superconductivity in non-centrosymmetric superconductors} (Gordon and Breach,
  1999)

\bibitem{maki:69}
K.~Maki, \emph{Superconductivity} (Marcel Dekker, 1969), Vol.~2, chap.~18, p.
  1035

\bibitem{liu:08}
B.~Liu, I.~Eremin, Phys. Rev. B \textbf{78}, 014518 (2008)

\bibitem{akbari:10}
A.~Akbari, I.~Eremin, P.~Thalmeier, Phys. Rev. B \textbf{81}, 014524 (2010)

\bibitem{yuan:06}
H.Q. Yuan, D.F. Agterberg, N.~Hayashi, P.~Badica, D.~Vandervelde, K.~Togano,
  M.~Sigrist, M.B. Salamon, Phys. Rev. Lett. \textbf{97}, 017006 (2006)

\end{thebibliography}
% etc \end{thebibliography}

\end{document}